# Near-Field Thermal Emission by Periodic Arrays

Sheila Edalatpour[†]

*Department of Mechanical Engineering, University of Maine, Orono, ME 04469, USA*

*Frontier Institute for Research in Sensor Technologies, University of Maine, Orono, ME 04469, USA*

## ABSTRACT

Near-field thermal emission can be engineered by using periodic arrays of sub-wavelength emitters. The array thermal emission is dependent on the shape, size, and materials properties of the individual elements as well as the period of the array. Designing periodic arrays with desired properties requires models that relate the array geometry and material properties to its near-field thermal emission. In this study, a periodic method is presented for modeling two-dimensional periodic arrays of sub-wavelength emitters. This technique only requires discretizing one period of the array, and thus is computationally beneficial. In this method, the energy density emitted by the array is expressed in terms of array's Green's functions. The array Green's functions are found using the discrete dipole approximation in a periodic manner by expressing a single point source as a series of periodic arrays of phase-shifted point sources. The presented method can be employed for modeling periodic arrays made of inhomogeneous and complex-shape emitters with non-uniform temperature distribution. The proposed technique is verified against the non-periodic thermal discrete-dipole-approximation simulations, and it is demonstrated that this method can serve as a versatile and reliable tool for studying near-field thermal emission by periodic arrays.

## I. INTRODUCTION

Thermal emission is in the near-field regime when the observation distance from the emitter is smaller than or comparable to the dominant thermal wavelength as determined using Wien's displacement law. Otherwise, thermal emission is said to be in the far-field regime. While far-field thermal emission is broadband, incoherent, unpolarized and limited by blackbody radiation, near-field thermal emission can be quasi-monochromatic, coherent, polarized and exceeds the

---

[†] Corresponding author. Tel.: +1 207 581 2375, Fax: +1 207 581 2379

E-mail address: sheila.edalatpour@maine.edu

blackbody limit by several orders of magnitude [1]. These properties of thermal near field are capitalized on for many promising applications in waste heat recovery [2–8], thermal rectification [9–15], nanoscale imaging [16–19], and nanomanufacturing [20–22]. Most of these applications require near-field properties that are not found among natural materials. Near-field thermal emission can be engineered by using periodic arrays of sub-wavelength emitters [23,24,33–36,25–32]. Designing periodic arrays with desired properties requires models that relate the geometry and material properties of the array to its near-field thermal emission. Analytical models do not exist for this purpose. As such, periodic arrays have been modeled using the effective medium theory (EMT) [23–26,28,31–33,35] or by employing numerical methods [27,30,44,45,36–43]. The EMT is an approximate method in which the array is modeled as a homogenous medium with effective dielectric properties. The validity of the EMT in the near-field regime, where the observation distance is in the same order of magnitude as the emitter sizes, is questionable. Numerical simulation of periodic arrays is done either by modeling an effective length of the array [27,30,36] or by exploiting the periodicity and modeling only a period of the array [37–45]. Modeling an effective length of the array, which usually comprises of several periods, can be computationally expensive. Particularly, a greater number of periods needs to be discretized as the observation distance increases. Furthermore, simulations should be repeated a few times to ensure that the number of periods selected for modeling results in a converged solution. It is very beneficial to have periodic numerical methods in which only a period of the array is discretized. So far, periodic modeling of near-field thermal emission is done for rectangular, triangular and ellipsoidal gratings [37–45]. One-dimensional rectangular gratings are modeled using the scattering approach [37–39], the finite-difference time-domain method (by applying the Bloch boundary conditions) [40], as well as the rigorous coupled wave analysis (RCWA) [41–44]. One-dimensional periodic arrays of triangular and ellipsoidal beams have been studied using the RCWA [43]. Near-fear heat transfer for a two-dimensional periodic array of rectangular gratings has also been calculated using a Wiener chaos formulation [45]. In this paper, we present a periodic method based on the discrete dipole approximation (DDA) [46,47] which can be used for modeling two-dimensional periodic arrays of complex-shape emitters. The periodic emitters can be inhomogeneous and have non-uniform temperature distribution. In this method, the energy density emitted by the array is expressed in terms of array's Green's functions that are the response of the array to illumination by a single point source. The array Green's functions are found in a periodic manner by expanding the single point source into a series of periodic arrays of phase-shifted point sources. The DDA is used for numerical simulations. This approach requires modeling only one period of the array and thus is computationally beneficial. Although the array is assumed to be periodic in two dimensions, the proposed method can easily be applied for modeling one-dimensional and three-dimensional periodic arrays.

This paper is structured as follows. The problem under consideration is described and formulated in Sections II and III, respectively. The proposed periodic technique is discussed in Section IV, and numerical examples are provided in Section V. Finally, the concluding remarks are presented in Section VI.

## II. DESCRIPTION OF THE PROBLEM

A schematic of the problem under consideration is shown in Fig. 1. A two-dimensional, infinite array of arbitrarily-shaped objects is periodic in $x$- and $y$-directions. The array has periods $L_x$ and $L_y$ along the $x$- and $y$-directions, respectively, and is submerged in the free space. The smallest building block of the array, which can consist of an arbitrary number of arbitrarily-shaped objects, is referred to as the unit cell. The replica of the unit cell along $x$- and $y$-directions are numbered using variables $p$ and $q$, respectively, where $p$ and $q$ vary from $-\infty$ to $\infty$. The unit cell is identified as the cell with $(p, q) = (0, 0)$. The array is at a temperature $T$ greater than absolute zero and thus emits thermal radiation in the free space. The objects are assumed to be non-magnetic, isotropic, in local thermodynamic equilibrium, and their dielectric response is described by a frequency-dependent dielectric function $\varepsilon(\omega) = \varepsilon'(\omega) + i\varepsilon''(\omega)$. The spectral energy density, $u$, emitted by the array at an observation point $\mathbf{r}_o$ in the free space is to be calculated.

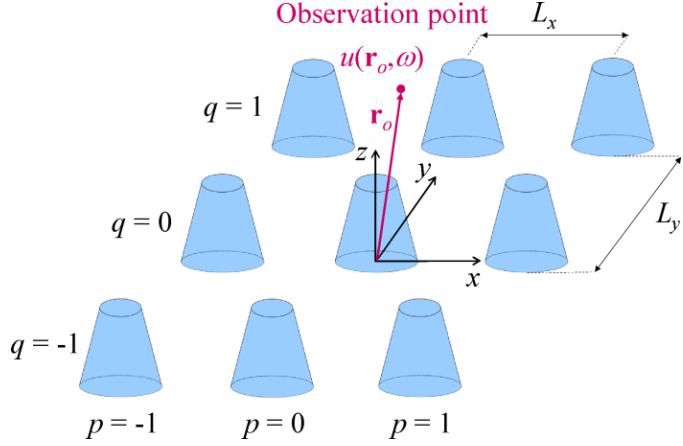

Figure 1. A schematic of the problem under consideration. A two-dimensional periodic array of arbitrarily-shaped objects with periods $L_x$ and $L_y$ in $x$- and $y$-directions, respectively, emits thermal radiation in the free space. The emitted energy density at the observation point $\mathbf{r}_o$, $u(\mathbf{r}_o,\omega)$, is desired.

## III. FORMULATION OF THE PROBLEM

The energy density at observation point $\mathbf{r}_o$ and angular frequency $\omega$ is given by [48]:

$$u(\mathbf{r}_o,\omega) = \frac{1}{2}\varepsilon_0 \text{Trace}\langle \mathbf{E}(\mathbf{r}_o,\omega)\otimes\mathbf{E}(\mathbf{r}_o,\omega)\rangle + \frac{1}{2}\mu_0 \text{Trace}\langle \mathbf{H}(\mathbf{r}_o,\omega)\otimes\mathbf{H}(\mathbf{r}_o,\omega)\rangle \tag{1}$$

where $\varepsilon_0$ and $\mu_0$ are the free space permittivity and permeability, respectively, $\mathbf{E}$ is the electric field, $\mathbf{H}$ is the magnetic field, $\otimes$ is the outer product and $\langle\ \rangle$ denotes the ensemble average. The electric field at point $\mathbf{r}_o$ can be obtained using the dyadic electric Green's function of the array $\mathbf{G}^E$ and the thermally fluctuating current $\mathbf{J}^{fl}$ as [49]:

$$\mathbf{E}(\mathbf{r}_o,\omega) = i\omega\mu_0 \int_V \mathbf{G}^E(\mathbf{r}_o,\mathbf{r}')\cdot \mathbf{J}^{fl}(\mathbf{r}',\omega)\,dV' \tag{2}$$

where $i$ is the imaginary unit number and the integral is performed over the volume of the array where the fluctuating current is non-zero. The dyadic electric Green's function $\mathbf{G}^E(\mathbf{r}_o,\mathbf{r}')$ relates the electric field at observation point $\mathbf{r}_o$ to the thermally fluctuating current at $\mathbf{r}'$ generating this electric field. The magnetic field at the observation point $\mathbf{r}_o$ can be obtained in a similar manner using the dyadic magnetic Green's function $\mathbf{G}^H$:

$$\mathbf{H}(\mathbf{r}_o,\omega) = \int_V \mathbf{G}^H(\mathbf{r}_o,\mathbf{r}') \cdot \mathbf{J}^{fl}(\mathbf{r}',\omega) dV' \qquad (3)$$

The ensemble average of the thermally fluctuating current $\mathbf{J}^{fl}$ is zero, while the ensemble average of its spatial correlation function is given by the fluctuation dissipation theorem [50,51]:

$$\langle \mathbf{J}^{fl}(\mathbf{r}',\omega) \otimes \mathbf{J}^{fl}(\mathbf{r}'',\omega) \rangle = \frac{4\omega\varepsilon_0\varepsilon''}{\pi} \Theta(\omega,T) \delta(\mathbf{r}'-\mathbf{r}'') \mathbf{I} \qquad (4)$$

In Eq. (4), $\varepsilon''$ is the imaginary part of the dielectric function of the objects, $\Theta(\omega,T) = \hbar\omega/[\exp(\hbar\omega/k_B T)-1]$, $\hbar$ and $k_B$ being the reduced Planck and Boltzmann constants, respectively, is the mean energy of an electromagnetic state [52], $\delta$ is the Dirac delta function, and $\mathbf{I}$ is the unit dyad. Substituting Eqs. (2) to (4) into Eq. (1), the energy density is written as:

$$u(\mathbf{r}_o,\omega) = \frac{2k_0^2}{\pi\omega} \int_V \varepsilon''\Theta(\omega,T) \text{Trace}\left[ k_0^2 \mathbf{G}^E(\mathbf{r}_o,\mathbf{r}') \otimes \mathbf{G}^E(\mathbf{r}_o,\mathbf{r}') + \mathbf{G}^H(\mathbf{r}_o,\mathbf{r}') \otimes \mathbf{G}^H(\mathbf{r}_o,\mathbf{r}') \right] dV' \qquad (5)$$

The electric and magnetic Green's functions of the array, $\mathbf{G}^E(\mathbf{r}_o,\mathbf{r}')$ and $\mathbf{G}^H(\mathbf{r}_o,\mathbf{r}')$, are needed for calculating the energy density using Eq. (5). Since the emitting array is made of isotropic and linear media, the reciprocity principle can be applied to this problem [53]. Based on the reciprocity principle, $\mathbf{G}^\gamma(\mathbf{r}_o,\mathbf{r}') = \left[\mathbf{G}^\gamma(\mathbf{r}',\mathbf{r}_o)\right]^T$ where $\gamma = E$ or $H$ [53,54]. As such, the energy density can equivalently be found using the Green's functions $\mathbf{G}^E(\mathbf{r}',\mathbf{r}_o)$ and $\mathbf{G}^H(\mathbf{r}',\mathbf{r}_o)$. As shown in Fig. 2, the electric Green's function is determined by calculating the electric field generated at $\mathbf{r}'$ due to radiation by a point source $\mathbf{J}(\mathbf{r})$ of magnitude $1/(i\omega\mu_0)$ located at $\mathbf{r}_o$ (i.e., $\mathbf{J}(\mathbf{r}) = \delta(\mathbf{r}-\mathbf{r}_o)/(i\omega\mu_0) \mathbf{I}$) [49]. In a similar manner and as shown in Fig. 2, the magnetic dyadic Green's function, $\mathbf{G}^H(\mathbf{r}',\mathbf{r}_o)$, is obtained by measuring the electric field at point $\mathbf{r}'$ due to a magnetic point source $\mathbf{M}(\mathbf{r})$ of unit magnitude radiating at $\mathbf{r}_o$ (i.e., $\mathbf{M}(\mathbf{r}) = \delta(\mathbf{r}-\mathbf{r}_o) \mathbf{I}$) [7]. This problem cannot be solved using the periodic DDA [55–57] which is employed for modeling light scattering by periodic arrays. In the periodic DDA, the array is illuminated by a planar incident field propagating at a given direction. However, in the problem shown in Fig. 2, the array is illuminated by a spherical wave generated due to radiation by the aperiodic point source. A periodic approach for solving this problem is presented in the next section.

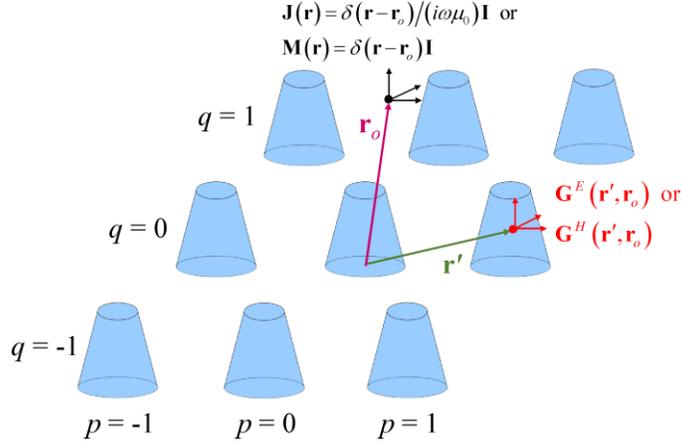

Figure 2. The dyadic electric (magnetic) Green's function for the array $\mathbf{G}^E(\mathbf{r}',\mathbf{r}_o)$ ($\mathbf{G}^H(\mathbf{r}',\mathbf{r}_o)$) is found by placing an electric (a magnetic) point source at $\mathbf{r}_o$ and measuring the electric field generated at $\mathbf{r}'$.

## IV. A PERIODIC APPROACH FOR CALCULATING ARRAY GREEN'S FUNCTIONS

The array Green's functions are obtained by calculating the electric field generated at point $\mathbf{r}'$ of the array due to radiation by a point source, represented by Dirac delta function, at the observation point $\mathbf{r}_o$. It is desired to solve this problem in a periodic manner. However, this problem is not periodic since the single point source illuminating the array is aperiodic. This aperiodic problem can be converted into a series of periodic problems by expressing the single point source as a periodic array of phase-shifted point sources.

### A. Periodic expansion of the Dirac delta function

The single point source emitting at $\mathbf{r}_o$ can be replaced by a periodic array of phase-shifted point sources with periods $L_x$ and $L_y$ using the fact that the Dirac delta function can be expanded as:

$$\delta(\mathbf{r}-\mathbf{r}_o) = \frac{L_x L_y}{(2\pi)^2} \int_{-\pi/L_x}^{\pi/L_x} \int_{-\pi/L_y}^{\pi/L_y} \sum_{p=-\infty}^{\infty} \sum_{q=-\infty}^{\infty} \delta\left(\mathbf{r}-\left[\mathbf{r}_o + pL_x\mathbf{x} + qL_y\mathbf{y}\right]\right) e^{i(pL_x k_x + qL_y k_y)} dk_y dk_x \qquad (6)$$

In Eq. (6), $\delta\left(\mathbf{r}-\left[\mathbf{r}_o + pL_x\mathbf{x} + qL_y\mathbf{y}\right]\right)$ represents the replica $(p, q)$ of the point source at $\mathbf{r}_o$ which is located at $\mathbf{r}_{opq} = \mathbf{r}_o + pL_x\mathbf{x} + qL_y\mathbf{y}$, $e^{i(pL_x k_x + qL_y k_y)}$ is the phase shift of the point source at $\mathbf{r}_{opq}$ relative to that located at $\mathbf{r}_o$, and $k_x$ and $k_y$ are the phasing gradients along the $x$- and $y$-directions, respectively. It should be noted that the phasing gradients are essentially mathematical wave vectors which are restricted to the periodicity of the Brillouin zone ($k_\beta$ between $-\pi/L_\beta$ and $\pi/L_\beta$, where $\beta = x$ and $y$). These mathematical wave vectors allow for the expansion of the delta function and by no means they represent a physical wavevector. Equation (6) holds true because when the phase-shifted point sources are integrated over the Brillouin zone, all of the point sources in the phase-shifted array integrate to zero except for the one located at $(p, q) = (0, 0)$. The proof of Eq. (6) is provided in the appendix. Using Eq. (6), the electric and magnetic point sources, $\mathbf{J}(\mathbf{r})$ and $\mathbf{M}(\mathbf{r})$, can be expressed as:

$$\mathbf{J}(\mathbf{r}) = \frac{L_x L_y}{(2\pi)^2} \int_{-\pi/L_x}^{\pi/L_x} \int_{-\pi/L_y}^{\pi/L_y} \sum_{p=-\infty}^{\infty} \sum_{q=-\infty}^{\infty} \frac{\delta\left(\mathbf{r}-\left[\mathbf{r}_o + pL_x\mathbf{x} + qL_y\mathbf{y}\right]\right)}{i\omega\mu_0} e^{i(pL_x k_x + qL_y k_y)} \mathbf{I} dk_y dk_x \qquad (7a)$$

$$\mathbf{M}(\mathbf{r}) = \frac{L_x L_y}{(2\pi)^2} \int_{-\pi/L_x}^{\pi/L_x} \int_{-\pi/L_y}^{\pi/L_y} \sum_{p=-\infty}^{\infty} \sum_{q=-\infty}^{\infty} \delta\left(\mathbf{r}-\left[\mathbf{r}_o + pL_x\mathbf{x} + qL_y\mathbf{y}\right]\right) e^{i(pL_x k_x + qL_y k_y)} \mathbf{I} dk_y dk_x \qquad (7b)$$

When $\mathbf{J}(\mathbf{r})$ and $\mathbf{M}(\mathbf{r})$ are expressed using Eqs. (7a) and (7b), a series of periodic problems such as the one shown in Fig. 3 is obtained. In these problems, a periodic array of objects is illuminated by a periodic array of phase-shifted point sources. Equations (7a) and (7b) show that the electric and magnetic point sources can be expressed as double integrations of Bloch waves with wavevectors limited to the first Brilloiun zone. The double summations within the integrals of Eqs. (7a) and (7b) show the expansion of these Bloch waves in terms of the reciprocal-lattice vectors. An analytical solution for the problem shown in Fig. 3 is not feasible and numerical solutions should be sought. A numerical solution based on the DDA is presented for this problem in the next section.

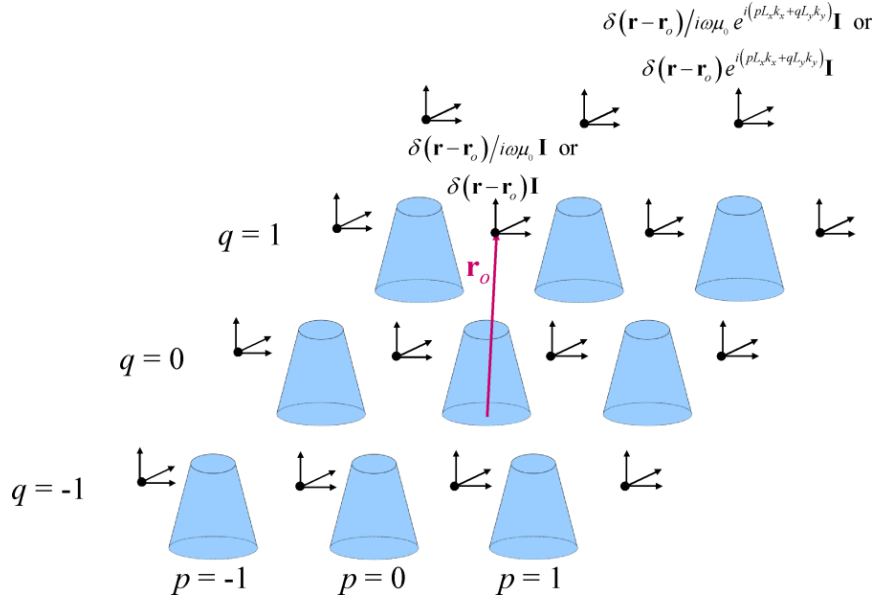

Figure 3. A single point source at the observation point $\mathbf{r}_o$ can be expressed as a series of periodic arrays of phase-shifted (relative to $\mathbf{r}_o$) point sources.

## B. Numerical solution of array Green's functions

The DDA, which is a volume discretization method [46,47,51], is used for calculating the Green's functions of the array. In this method, the objects are discretized into cubical sub-volumes with sizes much smaller than the thermal wavelength, the object sizes, the separation distance of the objects, and the distance between the observation point and the array. As such, the electric field can be assumed as uniform within the sub-volumes. It should be noted that the discretization size required for DDA simulations decreases as the refractive index of the emitters increases [51,58–61]. In this case, a volume discretization based on the Galerkin method of moments [61] can be computationally advantageous. By discretizing the volume-integral form of Maxwell's equations, the electric field in the sub-volumes can be written as [51]:

$$\frac{1}{\alpha_i} V_i \varepsilon_0 (\varepsilon_i - 1) \mathbf{E}_{imn} - k_0^2 \sum_{j=1}^{N} V_j (\varepsilon_j - 1) \sum_{p=-\infty}^{\infty} \sum_{\substack{q=-\infty \\ jpq \neq imn}}^{\infty} \mathbf{G}_{imn, jpq}^{0E} \cdot \mathbf{E}_{jpq} = \mathbf{E}_{imn}^{inc},$$

$$i = 1, 2, \ldots, N; \; m, n = 0, \pm 1, \pm 2, \ldots \quad (8)$$

where $\mathbf{E}_{imn}$ is the electric field in replica ($m$, $n$) of sub-volume $i$ in the unit cell (sub-volume $imn$) which is located at $\mathbf{r}_{imn} = \mathbf{r}_i + mL_x\hat{\mathbf{x}} + nL_y\hat{\mathbf{y}}$ with $\mathbf{r}_i$ being the position of sub-volume $i$, $\mathbf{G}^{0E}_{imn,jpq}$ is the free-space dyadic electric Green's function between sub-volumes $imn$ and $jpq$ [51,62], $\mathbf{E}^{inc}_{imn}$ is the electric field incident on sub-volume $imn$ due to radiation by the point-source arrays, $\alpha_i$ is the polarizability of sub-volume $i$ and its replica [51], and $k_0$ is the magnitude of the wavevector in the free space. The first summation in Eq. (8) runs over the $N$ sub-volumes located in the unit cell, while the second and third summations run over replica of the sub-volumes in the unit cell along the $x$- and $y$-directions, respectively. It should be noted that the double summation in Eq. (8) excludes the term corresponding to sub-volume $imn$, and it is assumed that the replica sub-volumes have the same dielectric function and volume as their counterpart in the unit cell.

For calculating the electric Green's function $\mathbf{G}^E_{imn,o}$, the incident field on the sub-volumes is due to radiation by the electric source $\mathbf{J}(\mathbf{r})$ given by Eq. (7a), and it can be calculated using the free-space electric Green's function as [49,63]:

$$\mathbf{E}^{inc}_{imn} = i\omega\mu_0 \int_V \mathbf{G}^{0E}(\mathbf{r}_{imn},\mathbf{r}) \cdot \mathbf{J}(\mathbf{r}) dV \tag{9}$$

Substituting for $\mathbf{J}(\mathbf{r})$ using Eq. (7a) and using the commutativity and associativity properties of the integral and summation, the incident field can be written as:

$$\mathbf{E}^{inc}_{imn} = \frac{L_x L_y}{(2\pi)^2} \int_{-\pi/L_x}^{\pi/L_x} \int_{-\pi/L_y}^{\pi/L_y} \sum_{p=-\infty}^{\infty} \sum_{q=-\infty}^{\infty} \int_V \mathbf{G}^{0E}(\mathbf{r}_{imn},\mathbf{r}) \delta\left(\mathbf{r} - \left[\mathbf{r}_o + pL_x\mathbf{x} + qL_y\mathbf{y}\right]\right) e^{i(pL_x k_x + qL_y k_y)} dV dk_y dk_x \tag{10}$$

which due to the fundamental property of the delta function reduces to:

$$\mathbf{E}^{inc}_{imn} = \frac{L_x L_y}{(2\pi)^2} \int_{-\pi/L_x}^{\pi/L_x} \int_{-\pi/L_y}^{\pi/L_y} \sum_{p=-\infty}^{\infty} \sum_{q=-\infty}^{\infty} \mathbf{G}^{0E}_{imn,opq} e^{i(pL_x k_x + qL_y k_y)} dk_y dk_x \tag{11}$$

For calculating the array magnetic Green's function $\mathbf{G}^H_{imn,o}$, the incident field is due to illumination by the magnetic current array $\mathbf{M}(\mathbf{r})$ as is given Eq. (7b). The incident electric field due to the magnetic source $\mathbf{M}(\mathbf{r})$ can be obtained in the same manner as for the electric current $\mathbf{J}(\mathbf{r})$, and it is given by:

$$\mathbf{E}_{imn}^{inc} = \frac{L_x L_y}{(2\pi)^2} \int_{-\pi/L_x}^{\pi/L_x} \int_{-\pi/L_y}^{\pi/L_y} \sum_{p=-\infty}^{\infty} \sum_{q=-\infty}^{\infty} \mathbf{G}_{imn,opq}^{0H} e^{i(pL_x k_x + qL_y k_y)} dk_y dk_x \qquad (12)$$

where $\mathbf{G}_{imn,opq}^{0H}$ is the free-space dyadic magnetic Green's function [48]. Equations (11) and (12) show that the incident electric field due to the electric and magnetic point sources can be written as integrals of Bloch waves with wavevector located in the first Brilloiun zone. The double summations within the integrals represent the expansion of the Bloch waves in the reciprocal-lattice domain. Substituting Eqs. (11) and (12) into Eq. (8) results in the following equation:

$$\frac{1}{\alpha_i} V_i \varepsilon_0 (\varepsilon_i - 1) \mathbf{G}_{imn,o}^{\gamma} - k_0^2 \sum_{j=1}^{N} V_j (\varepsilon_j - 1) \sum_{\substack{p=-\infty \\ jpq \neq imn}}^{\infty} \sum_{q=-\infty}^{\infty} \mathbf{G}_{imn,jpq}^{0E} \cdot \mathbf{G}_{jpq,o}^{\gamma}$$

$$= \frac{L_x L_y}{(2\pi)^2} \int_{-\pi/L_x}^{\pi/L_x} \int_{-\pi/L_y}^{\pi/L_y} \sum_{p=-\infty}^{\infty} \sum_{q=-\infty}^{\infty} \mathbf{G}_{imn,opq}^{0\gamma} e^{i(pL_x k_x + qL_y k_y)} dk_y dk_x ,$$

$$\gamma = E \text{ or } H; \; i = 1, 2, \ldots, N; \; m, n = 0, \pm 1, \pm 2, \ldots \qquad (13)$$

When Eq. (13) is written for all sub-volumes in the periodic array ($i = 1, 2, \ldots, N$; $m, n = 0, \pm 1, \pm 2, \ldots$), two self-consistent linear systems of equations are obtained which can be solved for $\mathbf{G}_{imn,o}^{E}$ and $\mathbf{G}_{imn,o}^{H}$. Since the system of equation (13) is linear, its solution can be written as the double integral of a wave-vector dependent Green's function, $\mathbf{g}_{imn,o}^{\gamma}(k_x, k_y)$, as:

$$\mathbf{G}_{imn,o}^{\gamma} = \frac{L_x L_y}{(2\pi)^2} \int_{-\pi/L_x}^{\pi/L_x} \int_{-\pi/L_y}^{\pi/L_y} \mathbf{g}_{imn,o}^{\gamma}(k_x, k_y) dk_y dk_x , \qquad \gamma = E \text{ or } H \qquad (14)$$

where $\mathbf{g}_{imn,o}^{\gamma}$ is the solution of the following equation:

$$\frac{1}{\alpha_i} V_i \varepsilon_0 (\varepsilon_i - 1) \mathbf{g}_{imn,o}^{\gamma} - k_0^2 \sum_{j=1}^{N} V_j (\varepsilon_j - 1) \sum_{\substack{p=-\infty \\ jpq \neq imn}}^{\infty} \sum_{q=-\infty}^{\infty} \mathbf{G}_{imn,jpq}^{0E} \cdot \mathbf{g}_{jpq,o}^{\gamma} = \sum_{p=-\infty}^{\infty} \sum_{q=-\infty}^{\infty} \mathbf{G}_{imn,opq}^{0\gamma} e^{i(pL_x k_x + qL_y k_y)} ,$$

$$\gamma = E \text{ or } H;\ i = 1, 2, \ldots, N;\ m, n = 0, \pm 1, \pm 2, \ldots \tag{15}$$

Equation (15) describes a periodic problem because the periodic array of the objects is illuminated by a periodic and phase-shifted incident field represented by the summation on the right-hand side of this equation. Since the problem is periodic and due to the translational symmetry of the free space Green's functions, the wave-vector dependent Green's function for sub-volume $jpq$, $\mathbf{g}^{\gamma}_{jpq,o}$, in Eq. (15) should be periodic and phase-shifted relative to that for sub-volume $j00$ located in the unit cell, $\mathbf{g}^{\gamma}_{j,o}$ [56]. As such, the wave-vector dependent Green's function of replica sub-volumes is related to that of their counterpart in the unit cell as:

$$\mathbf{g}^{\gamma}_{jpq,o} = \mathbf{g}^{\gamma}_{j,o} e^{i(pL_x k_x + qL_y k_y)}, \qquad p, q = 0, \pm 1, \pm 2, \ldots \tag{16}$$

It should be noted that the subscript 00 used for referring to the sub-volumes in the unit cell is dropped for simplicity. Equation (16) allows for solving the system of equations (15) only for the sub-volumes in the unit cell (i.e., for $i = 1, 2, \ldots, N$; $m = n = 0$). Substituting Eq. (16) into Eq. (15) and applying this equation to the sub-volumes in the unit cell results in:

$$\frac{1}{\alpha_i} V_i \varepsilon_0 (\varepsilon_i - 1) \mathbf{g}^{\gamma}_{i,o} - k_0^2 \sum_{j=1}^{N} V_j (\varepsilon_j - 1) \sum_{p=-\infty}^{\infty} \sum_{q=-\infty}^{\infty} \mathbf{G}^{0E}_{i,jpq} e^{i(pL_x k_x + qL_y k_y)} \cdot \mathbf{g}^{\gamma}_{j,o} = \sum_{p=-\infty}^{\infty} \sum_{q=-\infty}^{\infty} \mathbf{G}^{0\gamma}_{i,opq} e^{i(pL_x k_x + qL_y k_y)},$$

$$\gamma = E \text{ or } H;\ i = 1, 2, \ldots, N \tag{17}$$

where $\mathbf{g}^{\gamma}_{j,o}$ is taken out of the summations as it is independent of $p$ and $q$. The periodic free-space dyadic Green's function between two points $h$ and $l$ is defined as [64]:

$$\mathbf{G}^{0\gamma,P}_{h,l} = \sum_{p=-\infty}^{\infty} \sum_{q=-\infty}^{\infty} \mathbf{G}^{0\gamma}_{h,lpq} e^{i(pL_x k_x + qL_y k_y)}, \qquad \gamma = E \text{ or } H \tag{18}$$

where subscript $h$ indicates a sub-volume in the unit cell and $l$ refers to either a sub-volume in the unit cell or the observation point. Using the definition in Eq. (18), the system of equation (17) can be re-written in the following form:

$$\frac{1}{\alpha_i}V_i\varepsilon_0(\varepsilon_i-1)\mathbf{g}_{i,o}^{\gamma} - k_0^2\sum_{j=1}^{N}V_j(\varepsilon_j-1)\mathbf{G}_{i,j}^{0E,P}\cdot\mathbf{g}_{j,o}^{\gamma} = \mathbf{G}_{i,o}^{0\gamma,P}, \qquad \gamma = E \text{ or } H;\ i = 1, 2, \ldots, N \qquad (19)$$

Eq. (19) is dyadic, and 3 systems of equations of size $3N$ are obtained when it is applied to the sub-volumes in the unit cell ($i = 1, 2, \ldots, N$). The solution of these systems of equations provides the wave-vector dependent Green's functions for the sub-volumes located in the unit cell (i.e., $\mathbf{g}_{i,o}^{\gamma}$ where $i = 1, 2, \ldots, N$). The wave-vector dependent Green's function for replica sub-volumes ($\mathbf{g}_{imn,o}^{\gamma}$) is only phase-shifted relative to the sub-volumes in the unit cell ($\mathbf{g}_{i,o}^{\gamma}$), and it can be obtained using Eq. (16). Once the phase-dependent Green's functions $\mathbf{g}_{imn,o}^{\gamma}$ are found, the array Green's functions $\mathbf{G}_{imn,o}^{\gamma}$ can be calculated using Eq. (14). The energy density then can be found by using the discretized form of Eq. (5) and the array Green's functions $\mathbf{G}_{imn,o}^{\gamma}$ as:

$$u(\mathbf{r}_o,\omega) = \frac{2k_0^2 V_i}{\pi\omega}\sum_{i=1}^{N}\varepsilon_i''\Theta(\omega,T_i)\sum_{m=0}^{N_{kx}}\sum_{n=0}^{N_{ky}}\text{Trace}\left[k_0^2\mathbf{G}_{o,imn}^{E}\otimes\mathbf{G}_{o,imn}^{E} + \mathbf{G}_{o,imn}^{H}\otimes\mathbf{G}_{o,imn}^{H}\right] \qquad (20)$$

where $N_{kx}$ and $N_{ky}$ are the number of wavevectors selected for discretizing Brillouin zone along the $x$- and $y$-directions. The description of the periodic method for calculating energy density emitted by the periodic array is complete. The main steps in this technique can be summarized as follows.

1- The objects in the unit cell are discretized into $N$ cubical sub-volumes. The size of the sub-volumes should be much smaller than the thermal wavelength, the object sizes and the separation distances, such that the electric field can be assumed uniform in the sub-volumes.

2- Equation (19) is applied to the $N$ sub-volumes in the unit cell, and 3 systems of $3N$ equations are obtained. The solution of these systems of equations provides the wave-vector dependent Green's functions $\mathbf{g}_{i,o}^{\gamma}$ for the sub-volumes in the unit cell.

3- The wave-vector dependent Green's function for replica sub-volumes $\mathbf{g}_{imn,o}^{\gamma}$ is calculated using Eq. (16).

4- The array Green's function, $\mathbf{G}^{\gamma}_{imn,o}$, is found using the wave-vector dependent Green's functions, $\mathbf{g}^{\gamma}_{imn,o}$, and Eq. (14).

5- The energy density is calculated using the array Green's function, $\mathbf{G}^{\gamma}_{imn,o}$, and Eq. (19).

### C. Periodic free-space dyadic Green's functions

Calculating the periodic free-space dyadic Green's function defined in Eq. (18) requires evaluating two infinite summations. For near-field thermal radiation problems, these summations converge significantly faster in the reciprocal-lattice domain or using the Ewald method. Here we report the periodic dyadic Green's functions in the reciprocal-lattice domain. An Ewald representation can alternatively be utilized [64]. The periodic free-space scalar Green's function in the reciprocal domain is given by [64]:

$$G^{0,P}_{h,l} = \frac{i}{2L_x L_y} \sum_{p=-\infty}^{\infty} \sum_{q=-\infty}^{\infty} \frac{e^{ik_{z,pq}|z_h - z_l|}}{k_{z,pq}} e^{i\mathbf{K}_{pq} \cdot \boldsymbol{\rho}} \qquad (21)$$

where $\mathbf{K}_{pq} = (k_x + 2\pi p/L_x)\hat{\mathbf{x}} + (k_y + 2\pi q/L_y)\hat{\mathbf{y}}$ is the sum of the wave vector associated with the array of phase-shifted point sources and the wave vector of reciprocal lattice, $\boldsymbol{\rho} = (x_h - x_l)\hat{\mathbf{x}} + (y_h - y_l)\hat{\mathbf{y}}$ is the two-dimensional distance vector between points $h$ and $l$, and $k_{z,pq}$ is defined as:

$$k_{z,pq} = \sqrt{k_0^2 - K_{pq}^2}, \qquad \text{Im}[k_{z,pq}] \geq 0 \qquad (22)$$

The free-space electric dyadic Green's function can be determined by applying operator $\left[\mathbf{I} + \frac{1}{k_0^2}\nabla\nabla\right]$ on the scalar free-space Green's function given by Eq. (21) [49]. The result is:

$$\mathbf{G}^{0E,P}_{h,l} = \frac{i}{2L_x L_y} \sum_{p=-\infty}^{\infty} \sum_{q=-\infty}^{\infty} \left(\mathbf{I} - \frac{\mathbf{k}_{pq}\mathbf{k}_{pq}^T}{k_0^2}\right) \frac{e^{ik_{z,pq}|z_h - z_l|}}{k_{z,pq}} e^{i\mathbf{K}_{pq} \cdot \boldsymbol{\rho}} \qquad (23)$$

Where subscript $T$ means transpose, and $\mathbf{k}_{pq}$ is a wave vector defined as:

$$\mathbf{k}_{pq} = \mathbf{K}_{pq} + k_{z,pq} \frac{|z_h - z_l|}{z_h - z_l} \hat{\mathbf{z}} \tag{24}$$

The periodic free-space magnetic dyadic Green's function in the reciprocal domain can be found using the scalar periodic free-space Green's function as [49]:

$$\mathbf{G}_{h,l}^{0H,P} = \frac{1}{ik_0} \nabla \times \left( G_{h,l}^{0,P} \mathbf{I} \right) \tag{25}$$

which can be written as:

$$\mathbf{G}_{h,l}^{0H,P} = \frac{i}{2L_x L_y} \sum_{p=-\infty}^{\infty} \sum_{q=-\infty}^{\infty} \left( \frac{\mathbf{k}_{pq} \times \mathbf{I}}{k_0} \right) \frac{e^{ik_{z,pq}|z_h - z_l|}}{k_{z,pq}} e^{i\mathbf{K}_{pq} \cdot \boldsymbol{\rho}} \tag{26}$$

## V. NUMERICAL RESULTS

### A. Verification

The periodic technique presented in Section IV is verified against the thermal discrete dipole approximation (T-DDA) [51,65] simulations. Since the T-DDA is a non-periodic approach, an effective length of the array needs to be determined and modeled. The effective length is determined by increasing the array size until no significant change in the T-DDA results is observed. The periodic approach has been tested for two arrays. The first array, as shown in Fig. 4a, is made of silica nanospheres of diameter 10 nm separated by a distance of 40 nm in the $x$- and $y$-directions ($L_x = L_y = 50$ nm). The array emits at 400 K. The spectral energy density is calculated at two observation distances of 20 nm and 40 nm above the array along the perpendicular (to the array) axis of the nanospheres. The size of the particles is small compared to the wavelength, the period of the array, and the observation distance. As such, the nanospheres can be modeled as point dipoles using a single sub-volume. The effective length of the array required for the T-DDA simulations increases as the observation distance, $d$, increases. An effective length of 200 nm (equivalent to 25 periods of the array) is sufficient for calculating energy density at both observation distances of $d = 20$ nm and 40 nm. The $k_x$ and $k_y$ intervals (i.e., $[-\pi/L_x, \pi/L_x]$ and $[-\pi/L_y, \pi/L_y]$) in the periodic method are each discretized into 19 sub-intervals. The spectral energy density as calculated using the T-DDA (non-periodic approach) and the periodic approach is shown in Fig.

4b. The periodic and non-periodic simulations are in excellent agreement for both observation distances. There are two resonances in the spectrum of energy density at $\omega_{res,1} = 9.21 \times 10^{13}$ rad/s and $\omega_{res,2} = 2.13 \times 10^{14}$ rad/s. These resonances are due to the thermal excitation of localized surface phonons (LSPhs) of the silica nanospheres. Thermal emission by the nanospheres is proportional to the imaginary part of their polarizability $\alpha$ which is given by $\text{Im}[\alpha] = 9\varepsilon_0 V \varepsilon'' / |\varepsilon + 2|^2$. The LSPhs are excited at the frequencies for which the denominator of $\text{Im}[\alpha]$ vanishes, i.e., when $\varepsilon' \rightarrow -2$. This condition is satisfied at $\omega_{res,1}$ and $\omega_{res,2}$. The second array, which is shown in Fig. 4c, is made of silica nanoribbons of 50 nm height and 5 nm width. The nanoribbons are separated by 15 nm along the x- and y-directions such that $L_x = L_y = 20$ nm. The array emits at a temperature of 400 K. The spectral energy density is calculated at 15 nm and 30 nm above the array along the perpendicular (to the array) axis of the nanoribbons using the T-DDA and the periodic method. The nanoribbons need to be discretized into sub-volumes since they are large compared to the period of the array and the observation distance. As the size of the sub-volumes reduces, the accuracy of both methods increases until a converged solution is achieved for sufficiently small sub-volumes. A convergence analysis of the T-DDA can be found in Ref. [51]. For periodic simulations, the unit cell is discretized into 640 sub-volumes of size 1.25 nm, while the $k_x$ and $k_y$ intervals are each discretized into 23 equal sub-intervals. Modeling an effective length of 500 nm (equivalent to 625 periods of the array) is required for the T-DDA simulations. To reduce the computational time in the non-periodic T-DDA simulations, a non-uniform discretization comprising of 14720 sub-volumes of various sizes is used for discretizing the effective length of the array. In this non-uniform discretization, the size of the sub-volumes increases as their distance from observation point increases. The 9 periods of the array directly located underneath the observation point are each discretized into 640 sub-volumes of size 1.25 nm, the next 40 periods are each discretized into 80 sub-volumes of size 2.5 nm, and the remaining 576 periods are each discretized using 10 sub-volumes of size 5 nm. The results obtained using the two methods are shown in Fig. 4d, and they are in excellent agreement. The agreement between the T-DDA and the periodic approach, which are two different methods, confirms the validity of both approaches. The energy density has four resonances due to the excitation of LSPhs of the silica nanoribbons. These resonances are located at $\omega_{res,1} = 8.72 \times 10^{13}$ rad/s, $\omega_{res,2} = 9.24 \times 10^{13}$ rad/s, $\omega_{res,3} = 2.03 \times 10^{14}$ rad/s and $\omega_{res,4} = 2.14 \times 10^{14}$ rad/s for the observation distance of 15 nm. The LSPh resonances for the observation distance of 40 nm are observed at almost the same frequencies. The LSPhs are excited when the denominator of Im[$\alpha$] for nanoribbons vanishes. The polarizability of the nanoribbons can be estimated using that of a prolate spheroid with major and minor semi-axes equal to 25 nm and 2.5 nm, respectively. The imaginary part of the polarizability of a spheroid along the $j$ direction ($j = x$, $y$, and $z$) is given by Im[$\alpha_j$] = $\varepsilon_0 V \varepsilon'' / |1 + L_j(\varepsilon - 1)|^2$, where $L_j$ is the geometrical factor of the spheroid [66]. As such, LSPhs are observed at the frequencies for which $\varepsilon' \rightarrow (L_j - 1)/L_j$. For the nanoribbons, $L_x = L_y = 0.4899$ and $L_z = 0.0203$. Thus, the LSPhs along the major and minor axes are excited when $\varepsilon' \rightarrow$ -48.3 and $\varepsilon' \rightarrow$ -1.0, respectively. The first condition is satisfied at $\omega_{res,1}$ and $\omega_{res,3}$, while the second one holds true for $\omega_{res,2}$ and $\omega_{res,4}$.

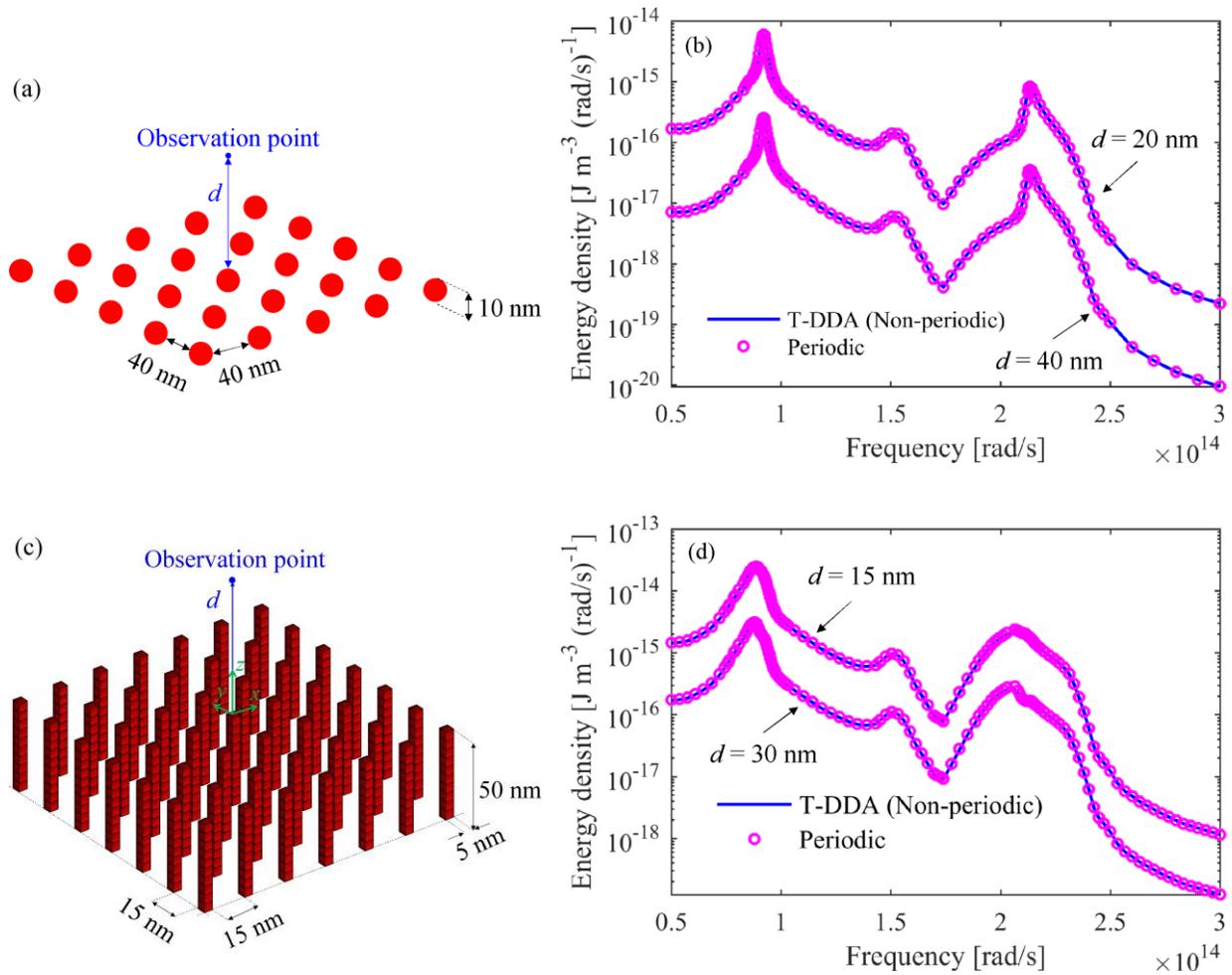

Figure 4. Schematics of periodic arrays of (a) nanospheres and (c) nanoribbons emitting at 400 K, and the spectral energy density emitted by (b) nanosphere and (d) nanoribbon arrays.

## B. Computational efficiency

In this sub-section, the computational resources (i.e., CPU time and memory) required for the periodic method are compared with those needed for the non-periodic T-DDA simulations. While only the unit cell of the array is discretized in the periodic method, the simulations should be repeated for a number of $k_x$ and $k_y$ values in the intervals $[-\pi/L_x, \pi/L_x]$ and $[-\pi/L_y, \pi/L_y]$, respectively. Additionally, infinite double summations should be evaluated for finding the periodic free-space Green's functions using Eqs. (25) and (26). Fortunately, these summations converge very rapidly in the reciprocal-lattice domain and they do not increase the computational time drastically. In the non-periodic T-DDA simulations, an effective length of the array comprising of several periods needs to be determined and modeled. As such, the number of sub-volumes in the T-DDA simulations is much greater than that in the periodic method. However, T-DDA simulations are not repeated for multiple values of $k_x$ and $k_y$. The CPU time in the T-DDA simulations increases with the number of sub-volumes, $N$, approximately as $N^3$ [51]. The number of sub-volumes can be written as $N = N_p N_{uc}$, where $N_p$ is the number of periods to be modeled and $N_{uc}$ is the number of sub-volumes used for discretizing the unit cell. Therefore, the CPU time in the T-DDA is proportional to $(N_p N_{uc})^3$. The memory required in the T-DDA for storing the complex-number elements of the interaction and dipole-moment correlation matrices with a double precision format is equal to $2.68 \times 10^{-7} (N_p N_{uc})^2$ GB. The number of periods of the array required for the T-DDA simulations, $N_p$, depends on the observation distance. As the observation distance increases, $N_p$ and consequently the CPU time (proportional to $N_p^3$) and the memory (proportional to $N_p^2$) required for the T-DDA simulations increase very rapidly. For this reason, modeling thermal emission at observation distances larger than a few periods of the array using the T-DDA becomes intractable. Excluding the time required for computing the periodic free-space Green's functions, the CPU time in the periodic method varies as $N_{kx} N_{ky} N_{uc}^2$, where $N_{kx}$ and $N_{ky}$ are the number of wavevectors along the $x$- and $y$-directions, respectively. The memory required in the periodic method for storing the Green's functions of the periodic array for various values of $k_x$ and $k_y$ is equal to $2.68 \times 10^{-7} N_{k_x} N_{k_y} N_{uc}^2$ GB. Based on the above discussion, it can roughly be concluded that when $N_{kx} N_{ky}$ is smaller than $N_p^3$ and $N_p^2$, the periodic method is advantageous with regard to CPU time and memory, respectively. This is usually the case especially when considering medium and large observation distances from the array.

As an example, the CPU time and memory required for modeling energy density emitted by the nanoribbon array in Fig. 4c using the T-DDA and the periodic method are reported in Table 1 for two observation distances of $d = 10$ nm ($d/L_x = 0.5$) and 30 nm ($d/L_x = 1.5$). The energy density is calculated at an angular frequency of $1.0 \times 10^{14}$ rad/s and a temperature of 400 K. The unit cell of the array is discretized into 640 sub-volumes of size 1.25 nm. In periodic simulations, the $k_x$ and $k_y$ intervals are each divided into 23 equal sub-intervals. In the T-DDA simulations, the number of periods of the array is increased until the energy density is within 1% of that predicted using the periodic method. For $d = 10$ nm, 25 periods of the array (equivalent to an effective length of 80 nm and 16000 sub-volumes of size 1.25 nm) are required to achieve a converged solution using the T-DDA. While both methods require approximately the same amount of memory, the periodic method is more than 58 times faster than the T-DDA. When the observation distance is increased to $d = 1.5L_x$, the CPU time and memory in the periodic method remain the same. However, modeling 289 periods of the array (equivalent to an effective length of 320 nm and 184960 sub-volumes of size 1.25 nm) is needed for the T-DDA simulations. Since modeling 184960 sub-volumes using the T-DDA requires significant computational resources, a non-uniform discretization scheme is employed for this case. In the non-uniform discretization, the 9 periods of the array directly located underneath the observation point are each discretized using 640 sub-volumes of size 1.25 nm, the next 40 periods are each discretized using 80 sub-volumes of size 2.5 nm, and the remaining 240 periods are each discretized using 10 sub-volume of size 5 nm. In total, the array is discretized into 11360 sub-volumes with non-uniform sizes. The CPU time for modeling this array is larger than that of the periodic method by more than 21 times, not to mention the additional time required for repeating simulations to ensure convergence and designing a non-uniform discretization. As the observation distance increases further, the T-DDA simulations become increasingly more challenging. It is also worth mentioning that the computations in the periodic method are highly parallelizable, since the array Green's functions can be calculated independently for each value of $k_x$ and $k_y$.

Table 1. The CPU time and memory used for modeling the energy density emitted by the nanoribbon array in Fig. 4c using the periodic and non-periodic (T-DDA) methods. The energy density is calculated at 1.0 rad/s and 400 K. In Table 1, U (NU) indicates that uniform (non-uniform) sub-volumes are used.

|  | $d/L_x = 0.5$ ($d = 10$ nm) | | | | $d/L_x = 1.5$ ($d = 30$ nm) | | | |
|---|---|---|---|---|---|---|---|---|
|  | $u \times 10^{14}$ [Jm$^{-3}$(rad/s)$^{-1}$] | $N$ | CPU time [s] | Memory [GB] | $u \times 10^{16}$ [Jm$^{-3}$(rad/s)$^{-1}$] | $N$ | CPU time [s] | Memory [GB] |
| Periodic | 1.653 | 640 U | 1326 | 63.2 | 3.654 | 640 U | 1348 | 63.2 |
| Non-periodic | 1.644 | 16000 U | 77913 | 68.6 | 3.623 | 11360 NU | 29076 | 34.6 |

## VI. CONCLUSIONS

Near-field thermal emission by periodic arrays was modeled using a periodic technique which only requires discretizing one period of the array. This technique is based on the DDA and expressing a single point source in terms of a series of periodic arrays of phase-shifted point sources. The near-field energy density emitted by periodic arrays of silica nanospheres and nanoribbons was modeled using the presented technique and by direct numerical modeling using the T-DDA. An excellent agreement existed between the two techniques which demonstrates the validity of the periodic technique. The proposed technique is efficient and versatile, and it can be used for modeling a wide variety of arrays comprised of complex-shape emitters. The emitters can be inhomogeneous with non-uniform temperature distribution. However, the inhomogeneity and the temperature distribution should be periodic.

## ACKNOWLEDGMENTS

This work is funded by National Science Foundation under Grant number CBET-1804360.

## APPENDIX: PERIODIC EXPANSION OF THE DIRAC DELTA FUNCTION

In this appendix, it is proved that the Dirac delta function can be expressed as a series of periodic arrays of phase-shifted delta functions with arbitrary periods $L_x$ and $L_y$. Mathematically, this is written as:

$$\delta(\mathbf{r}-\mathbf{r}_o) = \frac{L_x L_y}{(2\pi)^2} \int_{-\pi/L_x}^{\pi/L_x} \int_{-\pi/L_y}^{\pi/L_y} \sum_{p=-\infty}^{\infty} \sum_{q=-\infty}^{\infty} \delta\left(\mathbf{r}-\left[\mathbf{r}_o + pL_x\mathbf{x} + qL_y\mathbf{y}\right]\right) e^{i\left(pL_x k_x + qL_y k_y\right)} dk_y dk_x \quad (A.1)$$

Using the commutative property of integral and summation, the right-hand side of Eq. (A.1) can be written as:

$$\begin{aligned}
&\frac{L_x L_y}{(2\pi)^2} \int_{-\pi/L_x}^{\pi/L_x} \int_{-\pi/L_y}^{\pi/L_y} \sum_{p=-\infty}^{\infty} \sum_{q=-\infty}^{\infty} \delta\left(\mathbf{r}-\left[\mathbf{r}_o + pL_x\mathbf{x} + qL_y\mathbf{y}\right]\right) e^{i\left(pL_x k_x + qL_y k_y\right)} dk_y dk_x \\
&= \frac{L_x L_y}{(2\pi)^2} \sum_{p=-\infty}^{\infty} \sum_{q=-\infty}^{\infty} \delta\left(\mathbf{r}-\left[\mathbf{r}_o + pL_x\mathbf{x} + qL_y\mathbf{y}\right]\right) \int_{-\pi/L_x}^{\pi/L_x} e^{ipL_x k_x} \int_{-\pi/L_y}^{\pi/L_y} e^{iqL_y k_y} dk_y dk_x
\end{aligned} \quad (A.2)$$

where $\int_{-\pi/L_y}^{\pi/L_y} e^{iqL_y k_y} dk_y$ is zero unless $q = 0$. For $q = 0$, this integral equals $\frac{2\pi}{L_y}$. As such, Eq. (A.2) can be re-written as:

$$\begin{aligned}
&\frac{L_x L_y}{(2\pi)^2} \int_{-\pi/L_x}^{\pi/L_x} \int_{-\pi/L_y}^{\pi/L_y} \sum_{p=-\infty}^{\infty} \sum_{q=-\infty}^{\infty} \delta\left(\mathbf{r}-\left[\mathbf{r}_o + pL_x\mathbf{x} + qL_y\mathbf{y}\right]\right) e^{i\left(pL_x k_x + qL_y k_y\right)} dk_y dk_x \\
&= \frac{L_x}{2\pi} \sum_{p=-\infty}^{\infty} \delta\left(\mathbf{r}-\left[\mathbf{r}_o + pL_x\mathbf{x}\right]\right) \int_{-\pi/L_x}^{\pi/L_x} e^{ipL_x k_x} dk_x
\end{aligned} \quad (A.3)$$

In the same manner, $\int_{-\pi/L_x}^{\pi/L_x} e^{ipL_x k_x} dk_x$ is only non-zero when $p = 0$ in which case the integral equals $\frac{2\pi}{L_x}$. As such, the right-hand side of Eq. (A.3) reduces to $\delta(\mathbf{r}-\mathbf{r}_o)$.

**REFERENCES**

[1]   K. Joulain, J.-P. Mulet, F. Marquier, R. Carminati, and J.-J. Greffet, Surf. Sci. Rep. **57**, 59 (2005).

[2]   R. S. DiMatteo, P. Greiff, S. L. Finberg, K. A. Young-Waithe, H. K. H. Choy, M. M. Masaki, and C. G. Fonstad, Appl. Phys. Lett. **79**, 1894 (2001).

[3]     M. D. Whale and E. G. Cravalho, IEEE Trans. Energy Convers. **17**, 130 (2002).

[4]     M. Laroche, R. Carminati, and J. J. Greffet, J. Appl. Phys. **100**, (2006).

[5]     K. Park, S. Basu, W. P. King, and Z. M. Zhang, J. Quant. Spectrosc. Radiat. Transf. **109**, 305 (2008).

[6]     M. Francoeur, R. Vaillon, and M. P. Mengüç, IEEE Trans. Energy Convers. **26**, 686 (2011).

[7]     M. P. Bernardi, O. Dupré, E. Blandre, P.-O. Chapuis, R. Vaillon, and M. Francoeur, Sci. Rep. **5**, 11626 (2015).

[8]     A. Fiorino, L. Zhu, D. Thompson, R. Mittapally, P. Reddy, and E. Meyhofer, Nat. Nanotechnol. 1 (2018).

[9]     C. R. Otey, W. T. Lau, and S. Fan, Phys. Rev. Lett. **104**, 1 (2010).

[10]    S. Basu and M. Francoeur, Appl. Phys. Lett. **98**, 113106 (2011).

[11]    L. P. Wang and Z. M. Zhang, Nanoscale Microscale Thermophys. Eng. **17**, 337 (2013).

[12]    P. Ben-Abdallah and S. A. Biehs, Appl. Phys. Lett. **103**, 2011 (2013).

[13]    K. Ito, K. Nishikawa, H. Iizuka, and H. Toshiyoshi, Appl. Phys. Lett. **105**, 1 (2014).

[14]    K. Joulain, Y. Ezzahri, J. Drevillon, B. Rousseau, and D. De Sousa Meneses, Opt. Express **23**, A1388 (2015).

[15]    A. Ghanekar, Y. Tian, M. Ricci, S. Zhang, O. Gregory, and Y. Zheng, Opt. Express **26**, A209 (2018).

[16]    Y. De Wilde, F. Formanek, R. Carminati, B. Gralak, P.-A. Lemoine, K. Joulain, J.-P. Mulet, Y. Chen, and J.-J. Greffet, Nature **444**, 740 (2006).

[17]    A. C. Jones and M. B. Raschke, Nano Lett. **12**, 1475 (2012).

[18]    Y. Kajihara, K. Kosaka, and S. Komiyama, Rev. Sci. Instrum. **81**, 1 (2010).

[19]    S. Komiyama, in *43rd Int. Conf. Infrared, Millimeter, Terahertz Waves* (2018), pp. 1–2.

[20]    H. J. Mamin, Appl. Phys. Lett. **69**, 433 (1996).

[21]    K. Wilder, C. F. Quate, D. Adderton, R. Bernstein, and V. Elings, Appl. Phys. Lett. **73**, 2527 (1998).

[22] E. A. Hawes, J. T. Hastings, C. Crofcheck, and M. P. Mengüç, Opt. Lett. **33**, 1383 (2008).

[23] M. Francoeur, S. Basu, and S. J. Petersen, Opt. Express **19**, 18774 (2011).

[24] S. Basu and L. Wang, Appl. Phys. Lett. **102**, (2013).

[25] S. J. Petersen, S. Basu, B. Raeymaekers, and M. Francoeur, J. Quant. Spectrosc. Radiat. Transf. **129**, 277 (2013).

[26] X. L. Liu, R. Z. Zhang, and Z. M. Zhang, Appl. Phys. Lett. **103**, 98 (2013).

[27] B. Liu and S. Shen, Phys. Rev. B **87**, 115403 (2013).

[28] B. Liu, J. Shi, K. Liew, and S. Shen, Opt. Commun. **314**, 57 (2014).

[29] J. Shi, B. Liu, P. Li, L. Y. Ng, and S. Shen, Nano Lett. **15**, 1217 (2015).

[30] A. Didari and M. P. Mengüç, Opt. Express **23**, A547 (2015).

[31] A. Ghanekar, L. Lin, J. Su, H. Sun, and Y. Zheng, Opt. Express **23**, A1129 (2015).

[32] J. Y. Chang, S. Basu, and L. Wang, J. Appl. Phys. **117**, (2015).

[33] J. Y. Chang, Y. Yang, and L. Wang, Int. J. Heat Mass Transf. **87**, 237 (2015).

[34] K. Shi, F. Bao, and S. He, IEEE J. Quantum Electron. **54**, (2018).

[35] M. S. Mirmoosa, F. Rüting, I. S. Nefedov, and C. R. Simovski, J. Appl. Phys. **115**, (2014).

[36] A. D. Phan, T.-L. Phan, and L. M. Woods, J. Appl. Phys. **114**, 214306 (2013).

[37] G. Bimonte, Phys. Rev. A **80**, 042102 (2009).

[38] R. Guerout, J. Lussange, F. S. S. Rosa, J. Hugonin, D. A. R. Dalvit, J. Greffet, A. Lambrecht, and S. Reynaud, Phys. Rev. B **180301**, 1 (2012).

[39] J. Lussange, R. Guerout, F. S. S. Rosa, J. Greffet, A. Lambrecht, and S. Reynaud, Phys. Rev. B **085432**, (2012).

[40] A. W. Rodriguez, O. Ilic, P. Bermel, I. Celanovic, J. D. Joannopoulos, M. Soljačić, and S. G. Johnson, Phys. Rev. Lett. **107**, 114302 (2011).

[41] L. Wang, A. Haider, and Z. Zhang, J. Quant. Spectrosc. Radiat. Transf. **132**, 52 (2014).

[42] H. Chalabi, E. Hasman, and M. L. Brongersma, Phys. Rev. B **91**, 1 (2015).

[43] H. Chalabi, E. Hasman, and M. L. Brongersma, Phys. Rev. B **91**, 174304 (2015).

[44] H. Chalabi, A. Alù, and M. L. Brongersma, Phys. Rev. B **94**, 1 (2016).

[45] J. Li, B. Liu, and S. Shen, Phys. Rev. B **96**, 1 (2017).

[46] B. T. Draine, Astrophys. J. **333**, 848 (1988).

[47] M. A. Yurkin and a. G. Hoekstra, J. Quant. Spectrosc. Radiat. Transf. **106**, 558 (2007).

[48] J. D. Jackson, *Classical Electrodynamics*, 3rd ed. (John Wiley & Sons, New York, 1999).

[49] C.-T. Tai, *Dyadic Green Functions in Electromagnetic Theory*, 2nd ed. (IEEE Press, New York, 1994).

[50] Y. A. K. S. M. Rytov, Valerian Tatarskii, *Principles of Statistical Radiophysics 3: Elements of Random Fields* (Springer, New York, 1989).

[51] S. Edalatpour, M. Čuma, T. Trueax, R. Backman, and M. Francoeur, Phys. Rev. E **91**, 063307 (2015).

[52] Z. Zhang, *Micro/Nanoscale Heat Transfer* (McGraw-Hill Education, New York, 2007).

[53] C. A. Balanis, *Advanced Engineering Electromagnetics*, 2nd ed. (John Wiley & Sons, Inc., Hoboken, 2012).

[54] M. J. H. Ivar Stakgold, *Green's Functions and Boundary Value Problems*, 3rd ed. (John Wiley & Sons, New York, 2011).

[55] V. A. Markel, J. Mod. Opt. **40**, 2281 (1993).

[56] P. C. Chaumet, A. Rahmani, and G. W. Bryant, Phys. Rev. B **67**, 165404 (2003).

[57] B. T. Draine and P. J. Flatau, J. Opt. Soc. Am. A **25**, 2693 (2008).

[58] M. A. Yurkin and A. G. Hoekstra, J. Opt. Soc. Am. A **23**, 2578 (2006).

[59] M. A. Yurkin, V. P. Maltsev, and A. G. Hoekstra, J. Opt. Soc. Am. A, Opt. **23**, 2592 (2006).

[60] M. A. Yurkin and A. G. Hoekstra, J. Quant. Spectrosc. Radiat. Transf. **112**, 2234 (2011).

[61] A. G. Polimeridis, M. T. H. Reid, W. Jin, S. G. Johnson, J. K. White, and A. W. Rodriguez, Phys. Rev. B - Condens. Matter Mater. Phys. **92**, 1 (2015).

[62] L. Novotny and B. Hecht, *Principles of Nano-Optics*, 2nd ed. (Cambridge University Press, Cambridge, 2012).

[63] L. Tsang and J. A. Kong, *Scattering of Electromagnetic Waves: Advanced Topics* (John Wiley & Sons, Inc., New York, 2001).

[64] L. Tsang, J. A. Kong, K. Ding, and C. O. Ao, *Scattering of Electromagnetic Waves: Numerical Simulations* (John Wiley & Sons, Inc., New York, 2002).

[65] S. Edalatpour and M. Francoeur, J. Quant. Spectrosc. Radiat. Transf. **133**, 364 (2014).

[66] C. F. Bohren and D. R. Huffman, *Absorption and Scattering of Light by Small Particles* (Wiley-VCH, New York, 2007).